\documentclass{article}

\usepackage{soul}
\usepackage{url}
\usepackage[hidelinks]{hyperref}
\usepackage[utf8]{inputenc}
\usepackage[small]{caption}
\usepackage{graphicx}
\usepackage{amsmath}
\usepackage{booktabs}
\usepackage{amsthm}
\usepackage{latexsym}
\usepackage{amsfonts}
\usepackage{amssymb}
\usepackage{pifont}
\usepackage{natbib}
\setcitestyle{authoryear,open={(},close={)}}
\usepackage{algorithm}
\usepackage{algorithmicx}
\usepackage{algpseudocode}

\newcommand{\p}{{\rm P}}
\newcommand{\np}{{\rm NP}}

\newcommand{\elec}{\ensuremath{\mathcal{E}}}

\newtheorem{theorem}{Theorem}

\newtheorem{lemma}[theorem]{Lemma}
\newcommand\qedblob{\ding{113}}
\def\literalqed{{\ \nolinebreak\hfill\mbox{\qedblob\quad}}}

\newtheorem{observation}[theorem]{Observation}
\newtheorem{example}[theorem]{Example}

\newenvironment{proofs}{\noindent{\bf Proof.}\hspace*{1em}}{\literalqed\bigskip}

\showhyphens{}

\newcounter{int}
\makeatletter
\newcommand{\citen}[1] {\setcounter{int}{0}\@for\tmp:=#1\do{%
\ifnum \value{int}>0; \fi%
\setcounter{int}{1}%
\citeauthor{\tmp}~\cite{\tmp}}}
\makeatother

\newcommand{\prob}[3]{
\begin{description}
 \item[Name:] #1
 \item[Given:] #2
 \item[Question:] #3
\end{description}}

\begin{document}
\sloppy

\title{Selecting Voting Locations for Fun and Profit}

\author{Zack Fitzsimmons\footnote{Research done in part while on research visits to Ben-Gurion University and Rensselaer Polytechnic Institute.}\\
College of the Holy Cross\\
Worcester, MA USA\\
\and
Omer Lev\\
Ben-Gurion University\\
Beersheeba, Israel}

\date{March 15, 2020}

\maketitle

\begin{abstract}
While manipulative attacks on elections have been well-studied, only recently has attention turned to attacks that account for geographic information, which are extremely common in the real world.
The most well known in the media is gerrymandering, in which district border-lines are changed to increase a party's chance to win, but a different geographical manipulation involves influencing the election by selecting the location of polling places, as many people are not willing to go to any distance to vote. %
In this paper we initiate the study of this manipulation. We find that while it is easy to manipulate the selection of polling places on the line, it becomes difficult already on the plane or in the case of more than two candidates. Moreover, we show that for more than two candidates the problem is inapproximable. However, we find a few restricted cases on the plane where some algorithms perform well. Finally, we discuss how existing results for standard control actions hold in the geographic setting, consider additional control actions in the geographic setting, and suggest directions for future study.

\end{abstract}

\section{Introduction}

When faced with a set of different options, the use of voting as
a method to aggregate multiple agents' preferences on those options is thousands of years old; possibly as old as human society. As human societies grew, and the set of participants in the voting process grew with them, people quickly encountered a problem: How to deal with such a large set of people? Even Athenian democracy, which initially relied on a single decision making forum (the \emph{ecclesia}), was fairly quickly subdivided into subunits (\emph{demes}), based on geographic location. While many things have changed since then, the division of
political units by location is still very common. This is generally done in two ways:

\begin{enumerate}
\item Each subdivision makes its choice, and sends representatives to an aggregated assembly. This is the way elections are held in the US and in many Westminster type systems, in which subdivisions (e.g., UK constituencies, Canadian ridings, US districts) select a single representative, and send it to the national parliament.
\item Subdivisions are used for organizational purposes only---voting (or polling) places divide people according to where they live, but their vote is aggregated with other polling places in the same voting unit. In some countries (e.g., Israel), polling places are the only geographic division, while in others (e.g., US), polling places are the smallest geographic division, but others exist composed of units of polling places (counties, districts, states).
\end{enumerate}

The first case has been explored in the past few years in a series of papers (e.g., \cite{BLLZ16,BLSS18}), mainly discussing the potential for manipulation by drawing the subdivisions' borders (\emph{gerrymandering}). 
The second case of the subdivisions---used for organizational purposes only---has not, to the best of our knowledge, been significantly explored computationally.

The geographic problem in this case is not created by the voting rule, but by the location of the voters and the limited amount of effort each voter is willing or able to expend to cast their vote. Hence, the location of a polling place may have an effect on who votes there. Putting voting locations in convenient places encourages people to vote in the election, while locating them far away discourages participation. This form of control manipulation is already being implemented in various locations~\citep{Nic18}, and was recently a flashpoint in Georgia~\citep{Rei18}.

At its core, the problem is fairly straightforward: Voters are distributed in a geographic space, each having a maximal distance they are willing to travel in order to vote. Assuming $k$ polling places are to be located in a particular area, can putting them in particular places ensure a specified candidate's victory? While we shall use the geographic interpretation throughout this paper, we note that geographic distance is used as a proxy for difficulty of accessing a polling place, and other interpretations are possible. For example, in order to encourage students to submit faculty evaluations, the ``distance'' becomes how much effort
each student needs to put in depending on which platforms are used (e.g., paper forms, web, social networks, etc.).

Our results establish two parameters as key to the computational complexity of this problem.
The first is the dimension of the space, with the single dimension case easier than the plane and beyond. Note that solving the single dimensional problem is not useless as might be thought at first glance---not only can conceptual, nongeographic, settings be described by it, but even in political settings, voting locations might be located near a central highway or transportation route.

The second parameter is the number of parties/candidates, with the complexity changing once we leave the two-party system. Naturally, the two-party system is very common in many democracies, but even in settings where there are multiple parties, in many subunits (electoral districts, ridings, or constituencies) there are only two major competitive candidates, making the battle for electoral control mainly one in which two sides participate.

In this paper we \textbf{introduce a new model considering voter and polling place location}, and \textbf{formalize the associated control problem}. We show the complexity difference in the two-party case between the line and higher dimension, and between the two-party and multi-party case. Moreover, we show the \textbf{multi-party problem is inapproximable}. Finally, we discuss how \textbf{existing results translate to this model} and \textbf{novel results on new control problems} which are couched in real-world techniques.

\section{Related Work}

In this paper---as in geographic problems in general---we are particularly interested in control problems. That is, problems in which the design of the decision-making system is at play, and we wish to explore how changing the design can influence the decision outcome. 
The computational study of electoral control problems was introduced by \citet{bar-tov-tri:j:control} and 
includes natural scenarios such as control by adding candidates (modeling the addition of spoiler candidates to ensure a preferred outcome) and control by adding voters (modeling
get-out-the-vote drives). Later work by \citet{hem-hem-rot:j:destructive-control} considered the
so-called destructive case where the goal of the agent with control over the structure of the
election is to ensure that a despised candidate does not win, in contrast to the model
of constructive control studied by \citet{bar-tov-tri:j:control} where the goal of the agent
is to ensure that their preferred candidate wins. See the book chapter by \citet{fal-rot:b:handbook-comsoc-control-and-bribery} for
a recent survey of results on electoral control.

We use a distance-bound to model each voter's ability to vote at a polling place.
This is similar to the use of prices as used in priced electoral
control~\citep{mia-fal:j:priced-control}, and in the related problem of
bribery~\citep{fal-hem-hem:j:bribery}, where an agent sets the votes of a subcollection of the
voters to ensure a preferred outcome after meeting the price to change their vote.

Selecting a polling place in our model can be seen as adding the group of currently
nonparticipating voters within their distance-bound to that polling place to the election.
This idea of control by adding groups of voters was previously explored in different
ways by \citet{erd-hem-hem:c:natural-partition} and \citet{bul-che-fal-nie-tal:j:combinatorial-voter-control}.
Note that though we are adding ``groups'' of voters, this is quite different from the case of voters with weights.
A voter with weight $\omega$ can be thought of as a group
of $\omega$ voters with the same vote, but in our case a group of voters voting at a polling place may
have different votes and the ``weights'' at a polling place may be different depending on
if voters may be able to vote at a different polling place. The complexity of weighted control
was studied by \citet{fal-hem-hem:j:weighted-control}.

\citet{erd-hem-hem:c:natural-partition} consider a form of control
by adding voters where groups of voters must be added together. However,
in none of these models can the handling of overlap between groups be adapted to the geography-based groups we consider.

\citet{bul-che-fal-nie-tal:j:combinatorial-voter-control} introduced a model of
control by adding voters in which voters are grouped together (that is, one cannot add
a voter without adding some other voters) and explored different bundling functions to
define the groups. While this has some similarity with adding a polling place and getting along with it all voters which are within their distance-bound to its location, the entire conceptual framework does
not apply to our case.\footnote{The interesting properties they investigate in their models, such as leader-anonymous or follower-anonymous, have no relevance in our setting.}

There is also some relation to facility location problems, where the facility locations (which are the candidates) are from a fixed set (rather than everywhere), as explored in \citet{fel-fia-gol:c:voting-facility}. Note that many settings explored in this research direction tend to be on the one-dimensional line.

More closely related to our line of work are geographic manipulations of voting districts, commonly referred to as gerrymandering. This has been a topic of growing interest in the computational social choice community. The theoretical bounds on the influence of gerrymandering on the outcome of an election were established in \citet{BLLZ16}, and complexity results were shown in \citet{LLR17}, and expanded upon in \citet{CLR18}, focusing on graphs. We note that while \citet{LLR17} did not formally prove a greedy algorithm is an approximation algorithm for the gerrymandering problem, they did use it, de facto, as an algorithm to solve practical cases of gerrymandering (\citet{PPY17} suggested a different way to tackle gerrymandering, based on cake-cutting ideas). \citet{BLSS18} investigated how geographic spread affects gerrymandering ability, though those results are mostly empirical. Furthermore, there is a whole line of research on allowing agents to move between districts, a form of ``reverse gerrymandering,'' that has also been investigated in the past few years (see, \cite{BM12,BBCFNW15,LL19}).

\section{Preliminaries}

An election consists of a set of candidates $C$ and a collection of voters $V$ where each voter has a corresponding vote or preference order that strictly ranks the candidates in $C$ from most to least preferred. An election system \elec\ is a mapping from an election to a set of winning candidates. The best-known election system is plurality, in which each candidate gets a point from each voter that ranks them first in their preference order, and the candidates with the highest score win.

In our setting, in addition to their preferences over the candidates, all voters are located in a metric space ($\mathcal{M},d_{\mathcal{M}}$).\footnote{We stress that this is not putting the voters and candidates in an ``ideological'' metric space, {\`a} la~\citet{Sch08}---the candidates are not on the metric space, and voters are not single peaked in the sense of preferring candidates closer to them. This is literal physical space, which has nothing to do with the preferences.} Each voter $v\in V$ has an associated location $x_{v}\in\mathcal{M}$ and a nonnegative distance-bound $d_{v}\in\mathbb{R}$. We also have a set of potential polling places $L\subseteq\mathcal{M}$, each place $\ell\in L$ is defined using its location. A voter is able to vote only if there is a polling place $\ell$ for which $d_{\mathcal{M}}(x_{v},\ell)\leq d_{v}$.
Formally, the winner problem for an election system \elec\ in this setting:

\prob{\elec-Geographic-Winner}%
{A set of candidates $C$, a set of voters $V$ where each $v \in V$ has a location in a metric space $x_v \in \mathcal{M}$ (e.g., $\mathbb{R}^{2}$), a preference order $\succ_v$, and a distance-bound to vote $d_v$, a set of polling places $L \subseteq \mathcal{M}$ and a candidate $p \in C$.}%
{Is $p$ a winner using election system \elec\ when all voters $V$ within $d_v$ to a polling place in $L$ vote?}

To compute the geographic winner for an election system \elec, we must first
determine which voters are within their distance-bound to a polling place.
This can clearly be done in polynomial time. Thus each election system with
a polynomial-time winner problem has a polynomial-time geographic winner
problem.
More generally we can
state the following observation.

\begin{observation}
For every election system \elec\, the corresponding \elec-Geographic-Winner problem
 is in $\p^{\elec{\rm-winner}}$.
\end{observation}

\subsection{Electoral Control}

A natural model of electoral control in the geographic setting is to
consider how an election chair with control of the election can select polling places
to ensure their preferred outcome. It is realistic to assume that the election
chair is required to place at least a specified number of polling places, since
otherwise the election would be easily viewed as unfair, and so this is also included
in our model. We present the formal definition of the constructive case below (where
the goal of the election chair is to ensure a preferred candidate wins)~\cite{bar-tov-tri:j:control}, but we will also consider the destructive case (where the goal of the election chair is to ensure that a despised candidate does not win)~\cite{hem-hem-rot:j:destructive-control}.

\prob{\elec-Constructive Polling Place Control}%
{A set of candidates $C$, a set of voters $V$ where each $v \in V$ has a location in a metric space $x_v \in \mathcal{M}$, a preference order $\succ_v$, and a distance to vote $d_v$, a set of possible polling places in the plane $L \subseteq \mathcal{M}$, a preferred candidate $p \in C$, and a parameter $k$.}%
{Does there exist a set of polling places $L' \subseteq L$ such that $|L'| \ge k$ and $p$ is a winner of the geographic election $(C,V,L')$ using the election system \elec?}

\section{The Two-Party System}

In this section we will focus on settings of a two-party system, which is quite common across democratic countries. That is, we assume $|C|=2$.

We first consider a simpler version of our problem, where the voters and polling
places are on the real line instead of on the plane, i.e., $\mathcal{M}=\mathbb{R}$. For our voting scenario, this
can model situations such as the selection of polling places along a bus line in
a city, where the bus line's route can be viewed as a straight line.

\begin{theorem}
Constructive and Destructive Polling Place Control for plurality elections over two candidates with voters
located on the real line ($\mathcal{M}=\mathbb{R}$) is in \p.
\end{theorem}

\begin{proofs}
We consider the constructive case below. The constructive algorithm can be
easily adapted for the destructive case.

Let $C = \{p,r\}$, $V$, $L$, $p$, and $k$ be an instance of Polling Place Control
where the voters and the polling places are located on $\mathbb{R}$.
We will show that we can determine if $p$ can be made
a plurality winner in the geographic setting by selecting voting locations
using dynamic programming in polynomial time. %
Our dynamic programming algorithm works as follows.
We first order the polling places with respect to $\mathbb{R}$.
We construct the dynamic programming table so that 
$T[a, b, {\rm last}]$ is the maximum margin for $p$
(i.e., ${\rm score}(p) - {\rm score}(r)$) using $b$ of the first $a$ polling
places (ordered from left to right along $\mathbb{R}$), where ``${\rm last}$''
is the last polling place picked.
A solution to our problem is found if there is a nonnegative margin in the table
for $k$ polling places.
When filling out the table we can easily determine which voters
within their distance to a polling place under consideration are overlapping with
a polling place already picked with just the information for the last polling place
picked.~\end{proofs}

We now continue, and try to see if this result holds for more complex metric mechanisms.
As with related election problems with geographic constraints \citep{LLR17}, in general
the problem in the plane is \np-complete, as we present below.
We then discuss %
natural restrictions to this problem.

\begin{theorem}\label{thm:plane-np}
Constructive and Destructive Polling Place Control for plurality elections with voters located in the plane ($\mathcal{M}=\mathbb{R}^{2}$) is \np-complete for two candidates even when all voters have the same distance-bound.
\end{theorem}

\begin{proofs}
Membership of our problem in \np\ is easy to see.
\begin{figure}
\centering
\includegraphics[width=0.4\textwidth]{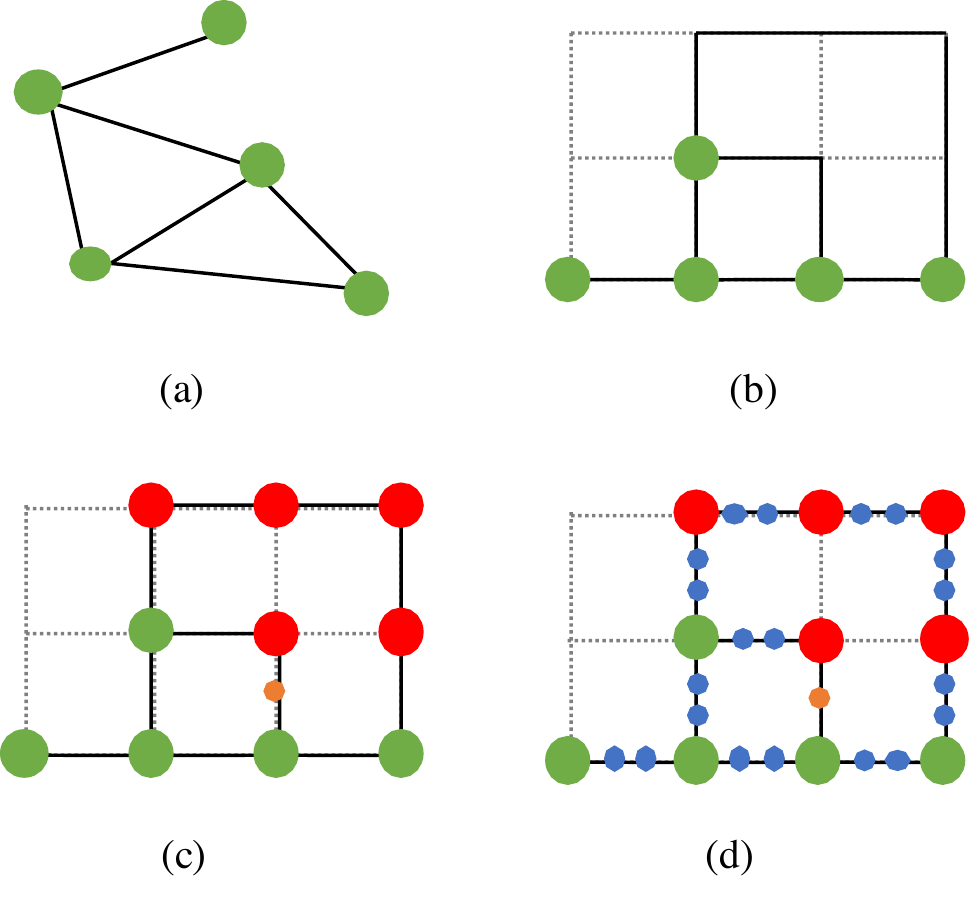}
\caption {The process to produce $\widehat{G}$ in the proof of Lemma~\ref{lem:res-plane-np}. (a) shows our initial graph; (b) shows a graph embedded onto the grid; %
(c) shows the addition of nodes at grid points, and a special orange node to make each edge have even number of additional vertices; (d) shows the added pairs of vertices dividing the larger edges.}
\label{graphConst}
\end{figure}

Our voting problem locates the voters in the plane, so the most straightforward
reduction will be to start with an appropriate planar \np-complete problem.

Planar Vertex Cover is \np-complete~\cite{gar-joh-sto:j:planar-node-cover}, even for cubic planar graphs (i.e., graphs where each vertex has degree at 
most three)~\citep{gar-joh:j:rectilinear-steiner}.

\prob{Cubic Planar Vertex Cover}%
{A Cubic (i.e., maximal degree is 3 or fewer) planar graph $G = (V, E)$ and an integer $k$.}%
{Does there exist a subset of vertices $V' \subseteq V$ such that $|V'| \le k$ and for each $(u,v) \in E$, $u \in V'$ or $v \in V'$?}

It will be useful to work with this problem embedded into the grid.
We will show that the following rectilinear version of Cubic Planar Vertex Cover
is \np-complete.\footnote{We mention that the some of our construction
follows the general contours
of a construction from~\citet{cha-hu:j:red-blue-setcover}, which shows a related
geometric set covering problem to be \np-hard.}

\prob{Restricted Rectilinear Cubic Planar Vertex Cover}%
{A  cubic planar graph $G = (V,E)$ embedded in a grid such that
all edges are on integer gridlines and of length 1 or 1.5, and an integer $k$.}%
{Does there exist a subset of vertices $V' \subseteq V$ such that $|V'| \le k$
and for each $(u,v) \in E$, $u \in V'$ or $v \in V'$?}

\begin{lemma}\label{lem:res-plane-np}
Restricted Rectilinear Cubic Planar Vertex Cover is \np-complete.
\end{lemma}

\begin{proofs}
Membership in \np\ is easy to see. 
Let $G = (V,E)$ and $k$ be an instance of Cubic Planar Vertex Cover.

\citet{val:j:vlsi-circuits} shows that a planar graph $(V, E)$ where each vertex has maximum degree four can be drawn in polynomial time in a $O(|V|) \times O(|V|)$ grid such that each vertex has integer coordinates and the edges are comprised of line segments along the integer gridlines, and the edges do not intersect with each other. We apply this construction to our graph and then subdivide the edges by adding vertices at each intersection of integer gridlines so that each edge in this new graph has length one. If an edge is not subdivided by an even number of vertices, we add an additional vertex at the midpoint between one of the original vertices of the edge and the closest vertex we added, so that overall, every original edge has been divided by adding an even number of vertices to the edge. Now all of the edges are length $1$ or $1/2$.

For each of the edges of length $1$, we add two additional vertices to subdivide the edge into three equal segments, i.e., one at $1/3$ and the other at $2/3$. Note that this maintains that each original edge has been divided by an even number of new vertices. We then rescale all of our edges by $3$ so that they are all of length $1.5$ or of length $1$. Denote this new graph $\widehat{G} = (\widehat{V}, \widehat{E})$ (see Figure~\ref{graphConst}).

When we subdivide an edge by adding two vertices to a given graph, the size of a minimum vertex cover increases by one.
Each edge in a graph has at least one of its vertices in a minimum vertex cover. When an edge is subdivided by
adding two vertices, it is ``replaced'' with three edges. One of the outer edges must be covered by a vertex in the original minimum vertex cover. To cover the edge between the two added vertices and the other outer edge,
one of the added vertices must be included in a minimum vertex cover of the updated graph.

It is straightforward to see that $G$ has a vertex cover of size at most $k$ if
and only if $\widehat{G}$ has a vertex cover of size $\hat{k} = k + 0.5t$,
with $t$ being the number of added vertices.~\end{proofs}

Having shown Lemma~\ref{lem:res-plane-np}, we will now put it to use.
Let $G = (V,E)$ and $k$ be an instance of Restricted Rectilinear Cubic Planar Vertex
Cover.
We now construct an instance of Constructive Polling Place Control for plurality elections (see Figure~\ref{edgeChanges} for an example of
a constructed edge).

\begin{figure}[t]
\centering
\includegraphics[width=0.75\linewidth]{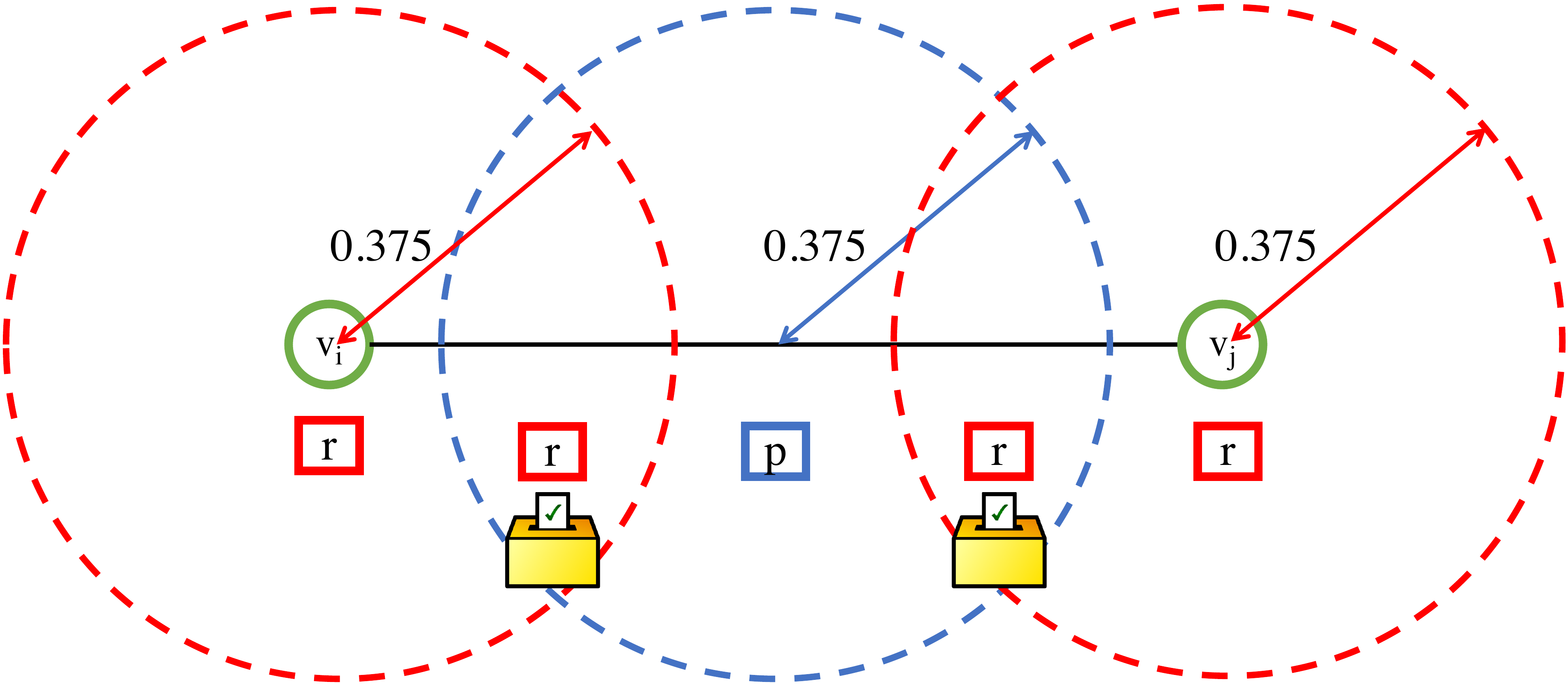}
\caption {The added voters and ballot boxes on an edge ($v_{i},v_{j}$). The circles denote the voters' distance-bounds. In order not to complicate the drawing further, the distance-bounds for the $r$ voters on the edges are not shown, but they are 0.375, thus extending past the $p$ voter on one side and a vertex on the other, but not as far as reaching another polling place.}
\label{edgeChanges}
\end{figure}

\begin{itemize}
 \item Let the set of candidates be $\{p,r\}$.
 \item For each vertex $v_i \in {V}$ create one voter for $r$ at that location with distance-bound $0.375$.
 \item For each edge $\{v_i, v_j\} \in {V}$ create one voter voting for $p$ located at the midpoint of the edge with distance-bound $0.375$.
 \item For each edge $\{v_i, v_j\} \in {V}$ add two polling places to $L$: one at the midpoint between the voter for $r$ at $v_i$ and the voter for $p$ at the midpoint, and the other at the midpoint between the voter for $r$ at $v_j$ and the voter for $p$
 \item At the location of every polling place, add an additional voter for $r$. This brings to a total of three voters located on each edge (without the vertices)---two supporting $r$ and one supporting $p$.
\item Add one additional polling place $\hat{q}$ at a distance strictly greater
 than $1.5$ from the constructed graph with an additional ${k}$ voters
 for $p$ located at that same location with distance-bound $0.375$.
\item Let the number of polling places to have be $|{E}|+1$.
\end{itemize}

We will now show that ${G}$ has a vertex cover of size
${k}$ if and only if there exists a subset of $|{E}| + 1$ polling
places to select such that $p$ is a plurality winner in the geographic setting.

Suppose that ${G}$ has a vertex cover of size ${k}$. Let this vertex
cover be $V'$. For each vertex $v \in V'$, select each polling place within a
distance of $0.375$ to $v$ if there is not already a polling place selected on
that edge. Since $V'$ is a vertex cover, we will have selected
$|{E}|$ polling places. It is easy to see that selecting these
polling places will result in $|{E}|$ voters for $p$ and
$|{E}| + {k}$ voters for $r$ to be within their
distance to a polling place. $p$ is made a winner by also selecting the
polling place $\hat{q}$.

For the converse, suppose that ${G}$ does not have a vertex cover of size
${k}$. Then we will show that there is no subset of at least $|{E}|+1$
polling places to add such that $p$ is a plurality winner in the geographic
setting. There are at most $|E|$ votes for $p$ among the polling places in $L'\setminus \hat{q}$, and if
we select two polling places on the same edge, the second only adds more votes for
$r$ (specifically either one or two additional votes for $r$), and no votes for $p$.
Assume that we have a subset $L'$ of at least $|{E}|+1$ polling places such that
$p$ is a plurality winner in the geographic setting. Then we know that 
$\hat{q}$ must be in $L'$, and the remaining polling places
selected can have at most ${k}$ more votes for $r$ than for $p$. Consider the remaining
polling places in $L' \setminus \hat{q}$. We know that at least $|{E}|$
polling places were selected, and they contribute at most ${k}$ votes for $r$ over $p$. However, the only polling places which do not add more $r$ votes than $p$ votes are those for which the $r$ voter located on a vertex already votes elsewhere. That is, the vertex voter overlaps for several polling places. If $p$ is a winner, this overlap must happen for at least $|E|-k$ polling places. That is, $|E|-k$ polling places did not change the balance between $p$ and $r$ and did not include the vertex $r$ voter. The $k$ that did include that voter are thus a vertex cover of $G$ of size $k$. This is a contradiction.

Notice that essentially the same reduction can be used for the destructive
case. There the goal would be to ensure that $r$ does not win.
And there would be $k+1$ votes for $p$ at $\hat{q}$. So then in
the case where there is a vertex cover, $p$ beats $r$, and in the case where
there is not a vertex cover, $r$ is a winner.~\end{proofs}

\subsection{Natural Restrictions to the Planar Case}

When we consider some natural restrictions
to the overlap between choice of polling places we have several tractable cases.
The simplest case is when no voter is able to vote at more than one of the polling
places under consideration. In this case it is easy to see that the greedy approach
of choosing the polling places in order of margin for $p$ is optimal when there
are two candidates.

\begin{theorem}\label{noOverlapTheorem}
Constructive and Destructive Polling Place Control for plurality elections over two candidates with voters
located in any metric space $\mathcal{M}$ is in \p\ when no voter $v \in V$ is within $d_v$ of more
than one polling place in $L$.
\end{theorem}

Notice that this result can be extended to instances where there is a fixed number
of polling places that share voters in common.

\begin{theorem}\label{boundedOverlapTheorem}
Constructive and Destructive Polling Place Control for plurality elections over two candidates with voters
located in a metric space $\mathcal{M}$ is in \p\ when the number of overlapping polling places is fixed.
\end{theorem}

\begin{proofs}
If there are $T$ overlapping polling places, this means there are at most $2^{T}$ different combinations of overlapping polling places that can be considered. For every $0\leq i\leq T$, one finds the best $k-i$ polling places from those that don't overlap any others (in \p\ from Theorem \ref{noOverlapTheorem}). Then it looks for the set of $i$ polling places from those that overlap, and see what set optimizes the outcome. Once all $T+1$ possibilities has been examined, chose the best one.~\end{proofs}

It is natural to wonder if this result can be pushed any further. For example,
for a fixed parameter $\ell$, each voter is within their distance to at most $\ell$
polling places. However, the construction from the proof of Theorem~\ref{thm:plane-np}
shows hardness for the case of $\ell = 3$. This leaves open the
question of what happens for the case of $\ell = 2$. In Example~\ref{greedyExample}
below we show that the obvious greedy approach will not solve this case.

\begin{example}\label{greedyExample}
Examine this instance of polling place control.
Let $C = \{p, r\}$ be the set of candidates, let the set of polling places
be $L = \{A,B,C\}$ with $A$ located at $(1,1)$, $B$ located at $(2,2)$,
and $C$ located at $(2.5,1)$. Let there be the following voters all with
distance $1$.

20 voters for $p$ at $(1.5,0.5)$ that can vote at $A$ or $C$.

30 voters for $p$ at $(1.5,1.5)$ that can vote at $A$ or $B$.

35 voters for $r$ at $(2.5,1.5)$ that can vote at $B$ or $C$.

51 voters for $r$ at $(1,1)$ that can vote at $A$.

5 voters for $r$ at $(2,2)$ that can vote at $B$.

\noindent
Let $k = 2$ and the preferred candidate be $p$ (see Figure~\ref{exampleGraphic}).

\begin{figure}
\centering
\includegraphics[width=0.5\linewidth]{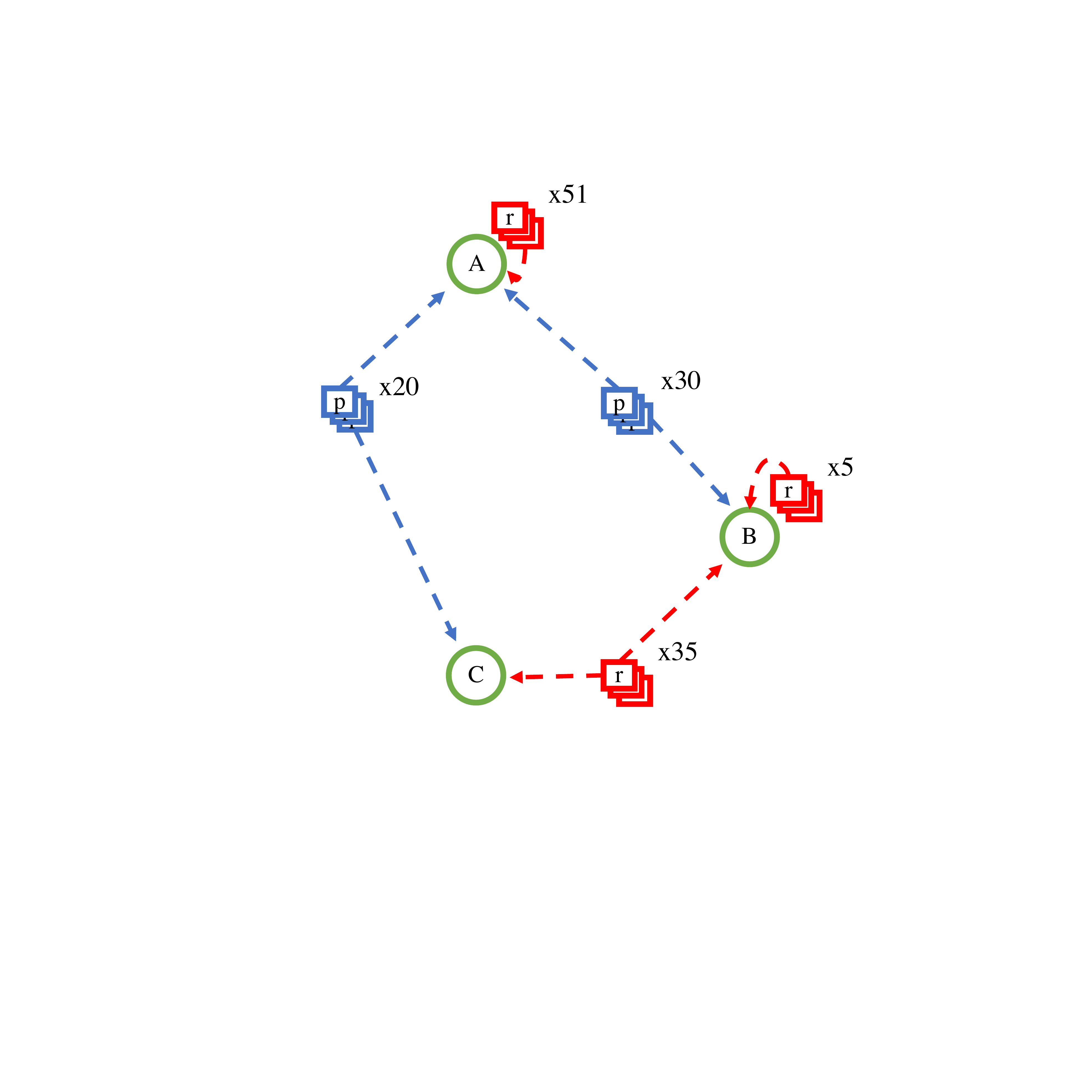}
\caption {The placement of voters in Example~\ref{greedyExample}. The dotted lines indicate where each set of voters can vote.}
\label{exampleGraphic}
\end{figure}

The margin for $p$ at $A$ is $-1$, at $B$ is $-10$, and at $C$ is $-15$. So the
greedy approach will choose $A$. After choosing $A$, the margin for $p$ at
$B$ is $-40$ and at $C$ is $-35$. And this algorithm would return that there
is no way to allocate at least 2 polling places such that $p$ wins.
However, if we instead choose the polling places $B$ and $C$, $p$ wins.
\end{example}

This greedy approach fails to find a solution when one exists due to the
overlap of voters between polling places, and so it cannot even be used as an approximation.

\section{Multi-Party System}

Unlike in the case of different distance-bounds for voters, when we move to
the case of more than two candidates, polling place control for plurality elections
becomes \np-complete even when voters can vote at most at one location.

When voters can vote at most at one location, %
the mapping to the metric space is trivial and we are instead left with considering how to add groups of voters to an initially empty set of voters to ensure that a preferred candidate wins. We show the following hardness result for the case of an unbounded number of candidates where the voting locations do not serve any of the same voters as another %
(that is, making the problem easier, as can be seen in the two-party case above).

\begin{theorem}\label{thm:hardness-multiple}
Constructive Polling Place Control for plurality elections is \np-complete for multiple
candidates even when voters can vote at most at one location and $\mathcal{M}=~\mathbb{R}$.
\end{theorem}

\begin{proofs}
Membership in \np\ is easy to see. We show \np-hardness by a reduction from
from Exact Cover by 3-Sets~\citep{kar:b:reducibilities}, defined below.

\prob{Exact Cover by 3-Sets}%
{A set $B = \{b_1, \ldots, b_{3k}\}$, and a collection $\mathcal{S} = \{S_1, \ldots, S_n\}$
of three-element subsets of $B$.}%
{Does there exist a subcollection $\mathcal{S}'$ of $\mathcal{S}$ such that every element of $B$
occurs in exactly one member of $\mathcal{S}'$?}

Given an instance of Exact Cover by 3-Sets, $B = \{b_1, \ldots, b_{3k}\}$ and $\mathcal{S} = \{S_1, \ldots, S_n\}$ such that each $S_i = \{b_{i1}, b_{i2}, b_{i3}\}$, we construct the following instance
of polling place control.

Let the set of candidates be $C = B \cup \{p\}$. In this construction, all voters will have
distance-bound $1/2$.
For each $S_i \in \mathcal{S}$, create a polling place at  location $(i)$ with $k-2$ votes for
each $b \in B$, with an additional $k$ votes for $b_{i1}$, $b_{i2}$, and $b_{i3}$ (so each of
these candidates has $2k-2$ voters at this polling place), and $k-1$ votes for $p$.
Let the preferred candidate be $p$, and let the number of polling places to be selected be $k$.

Suppose there exists an exact cover $\mathcal{S}'$. For each $S_i \in \mathcal{S}'$, add the
polling place at location $(i)$. Then each candidate $b_i$ has score exactly $(k-2)k+k = k^2 - k$ and $p$ has score $(k-1)k = k^2 - k$, and $p$ wins.

If there is a way to make $p$ win by choosing at least $k$ polling places then there must exist 
an exact cover. $p$ gets $k^2 - k$ points from any subset of $k$ polling places, and so the polling places selected must give each $b_i$ candidate at most $k^2 - k$ votes. Thus exactly $k$ of the polling places were selected and these $k$ polling places must correspond to an exact cover, otherwise some candidate $b_i$ has score greater than $k^2-k$.~\end{proofs}
\paragraph{Inapproximability}
Consider the optimization version of polling place control where we seek to maximize the number of polling places
selected. This problem exhibits nonmonotonicity as an optimization problem, i.e., for a given
set of polling places $L$ that ensures a given candidate $p$ wins it is not always the case that $p$
wins when a subset of $L$ was selected. More generally, for a given optimal solution
to this maximization problem, it is not the case that a worse solution is always valid.
And we show the following inapproximability result.

\begin{theorem}
It is NP-hard to approximate the number of polling places to select so that $p$
is a plurality winner in the geographic setting by any factor for the case of multiple
candidates even when voters can vote at most at one location and $\mathcal{M}=\mathbb{R}$.
\end{theorem}

The above proof follows from adapting the \np-hardness reduction from the proof
of Theorem~\ref{thm:hardness-multiple} to show that it is \np-hard to determine
if a given candidate $p$ is a plurality winner in the geographic setting by adding
a nonempty set of polling places, i.e, is it optimal to select 0 or more than 0
polling places to ensure $p$ wins. This implies that it is \np-hard to approximate
the optimization version of polling place control for plurality elections in this
multicandidate setting. We mention that a similar approach to showing inapproximability
was used by~\citet{car-cov-fel-hom-kak-kar-pro-ros:j:dodgson} to prove the inapproximability of Young score, another voting problem that exhibits nonmonotonicity as an optimization problem.

\subsection{Standard Control Actions in the Geographic Setting}

Our focus has been to explore how an election chair with control over the selection
of voting locations can ensure their preferred outcome in a geographic election.
However, as previously mentioned, this geographic model for elections is quite
natural and standard models of control (and many other election problems) can be
considered in this setting.

It is easy to see that \np-hardness results from the standard setting for control
by adding/deleting/partitioning candidates/voters in both the constructive and the destructive
cases are inherited to the corresponding actions
of control by adding/deleting/partitioning candidates/voters for geographic elections. First, we show that the existing problems' results extend to our new domain. %
For clarity
we formally define the problem of constructive control by deleting voters introduced
by \citet{bar-tov-tri:j:control}. Recall that the corresponding
destructive cases were introduced by \citet{hem-hem-rot:j:destructive-control}.\footnote{Both~\citet{bar-tov-tri:j:control} and~\citet{hem-hem-rot:j:destructive-control} use the so-called unique-winner model where the goal of the chair is to ensure that their preferred candidate is (is not) the unique winner, while we consider the now more commonly studied nonunique winner model. However, this section's results hold for both models.}
\prob{\elec-Constructive Control by Deleting Voters}%
{An election $(C,V)$, a delete limit $k$, and a preferred candidate $p \in C$.}%
{Does there exist a subcollection of voters $V' \subset V$ such that $|V'| \le k$
and $p$ is a winner of the election $(C, V \setminus V')$ using election system
\elec?}

The variant of the above control problem in the geographic setting replaces the
standard form of an election $(C,V)$ with our geographic variant $(C,V,L)$. Notice
that the polling places in the election are given, we are {\em not} selecting a set
of polling places for these variants.

\begin{observation}\label{obs:control-inheritance}
Each standard constructive and destructive control action
polynomial-time many one reduces
to the corresponding control action in the geographic setting.
\end{observation}

\begin{proofs} %
We consider the case of Constructive Control by Deleting Voters. The other cases follow from
essentially the same approach.

Given an instance of \elec-Constructive Control by Deleting Voters, $(C,V)$, $k$, and $p$,
we construct an instance of \elec-Constructive Control by Deleting Voters in the geographic setting.

Let the candidate set $C$, delete limit $k$, and preferred candidate $p$ remain the same.
For each voter $v_i \in V$, construct a voter
at $(0,i)$ with distance-bound $0.5$ and add a polling place at $(0,i)$ to $L$.

It is straightforward to see that if there exists a way to delete up to $k$ voters
such that $p$ is a winner, deleting these same voters in the constructed
geographic instance makes $p$ a winner in that setting and vice versa.
\end{proofs}%

As a result of Observation~\ref{obs:control-inheritance}, for a given election
system, if a standard control action is \np-hard, it remains \np-hard in the
geographic setting. At first glance it may seem that the structure of the geographic
setting may be able to realize an increase in the complexity of polynomial-time
control actions. However, this is not the case.

\begin{observation}
Each standard constructive and destructive control action in the geographic setting
polynomial-time many one reduces
to the corresponding standard control action.
\end{observation}

\begin{proofs}%
We consider the case of Constructive Control by Deleting Voters in the geographic
setting. The other cases follow from essentially the same approach.

Given an instance of \elec-Constructive Control by Deleting Voters in the geographic setting,
$(C,V,L)$, $k$, and $p$, we construct an instance of \elec-Constructive Control by
Deleting Voters.

Let the candidate set $C$, delete limit $k$, and preferred candidate $p$ remain
the same. For each $v \in V$ include $v$ in the constructed election if there exists
a polling place $\ell \in L$ such that
$d_{\mathcal{M}}(x_v, \ell) \le d_v$ (i.e., the voter $v$ can vote at the
polling place $\ell$).
It is easy to see that this reduction holds.
\end{proofs}%

\subsection{Additional Control Actions}

We can also define additional new natural models for control in the geographic setting. For example, consider the scenario where an agent can change the distance-bound for a subcollection of the voters to ensure their preferred outcome. We wish to stress that while in regular voting setting one might scoff at directly bribing voters as a rare occurrence (except as campaigning), in our setting, changing the distance bound for voters is easily done with perfectly legal means---organize a ride, or a bus, or any other form of transport to help voters reach their polling places. For preventing voter to vote, a candidate can cause traffic delays with rallies (or politically-devised traffic jams~\citep{Zer15}): 

\prob{\elec-Constructive Control by Distance-Bound Change}%
{A set of candidates $C$, a set of voters $V$ where each $v \in V$ has a location in a metric space $x_v \in \mathcal{M}$ (e.g., the plane, $\mathbb{R}^{2}$), a preference order $\succ_v$, and a distance-bound to vote $d_v$, a set of polling places $L \subseteq \mathcal{M}$ and a candidate $p \in C$,
and a budget $k$}%
{Do there exist subcollections of voters $V' \subseteq V$ and $W' \subseteq V$ with
$V' \cap W' = \emptyset$ such that
\scriptsize
\begin{equation*}
\begin{split}
\left(\sum_{v_i \in V'}{(d_{v_i} - \min(d_{v_{i}},d_{\mathcal{M}}(x_{v_i}, L_{v_i})) + \epsilon)}\right)+&\\
\left(\sum_{v_i \in W'}{(\max(d_{v_{i}},d_{\mathcal{M}}(x_{v_i},L_{v_i}))-d_{v_{i}})}\right) &\le k
\end{split}
\end{equation*}
\normalsize
for some $\epsilon>0$.\footnote{Practically, it is as small as one wishes. It is simply needed to make the distance of voter ${v_{i}}$ just a bit smaller than will make it able to reach its ballot box.} $L_{v_i}$ is the
location of the polling place nearest $v_i$ and $p$ is a winner using election system
$\elec$ of the geographic election $(C,\widehat{V},L)$ where $\widehat{V}$
consists of the voters from $V$ with updated distance-bounds for voters in $V'$ and $W'$?}

In the problem above, the difference between a voter's distance-bound and the distance
to the closest polling place can be viewed as their ``price'' for participating in
the election. We can show this problem to be in \p\ for plurality elections by
a reduction to multimode priced control by adding voters and deleting voters introduced by \citet{mia-fal:j:priced-control}.

\prob{\elec-Priced Constructive Control by Adding Voters and Deleting Voters}%
{An election $(C,V)$, a set of unregistered voters $W$, cost function $cost$ for each
voter in $V \cup W$, a candidate $p \in C$, and a budget $k$.}%
{Do there exist subcollections $V' \subseteq V$ and $W' \subseteq W$ such that
$cost(V' \cup W') \le k$
and $p$ is a winner of the election $(C, (V \setminus V') \cup W')$
using election system \elec?}

\begin{theorem}
Constructive Control by Distance-Bound Change for plurality elections is in \p.
\end{theorem}
\begin{proofs}
Let $(C,V,L)$, $k$, and $p$ be an instance of Control by Distance-Bound Change for
plurality elections. We construct the following instance of 
Priced Constructive Control by Adding Voters and Deleting Voters for plurality elections.

Since the algorithm developed by \citet{mia-fal:j:priced-control} uses only integer or rational numbers, we also need to convert our real
numbers into those. We take $\epsilon>0$ such that $2\epsilon$ is smaller than the amount any voter's distance-bound must change
to change from voting to not voting and vice versa, excluding voters within exactly their distance-bound to a polling place
(i.e., $2\epsilon < \min\{|d_{v}-d_{\mathcal{M}}(x_v,\ell)|\ |\ v\in V, \ell \in L, d_{\mathcal{M}}(x_v,\ell) \neq d_{v}\}$).\footnote{The algorithm
from~\citet{mia-fal:j:priced-control} is for the unique-winner winner model, but is easily adapted for the nonunique winner model.}

\begin{itemize}
    \item Let the set of candidates $C$ and preferred candidate $p$
    remain the same.
    \item We take as our budget $\hat{k}$ a rational number that is in the range $(k+\epsilon,k+\frac{|V|+1}{|V|}\epsilon)$.
    \item Let the set of registered voters $\widehat{V}$ consist of each voter
        $v \in V$ such that  there exists an $\ell \in L$ with 
        $d_{\mathcal{M}}(x_v,\ell) \le d_v$ (i.e., they are within their distance-bound to
        at least one polling place).
        If there exists an $\ell \in L$ such that $x_v = \ell$, set $cost(v) = \hat{k}+1$.\footnote{These voters
        are located exactly at a polling place and will vote even if their distance-bound is set to 0. Therefore
        we set the price of deleting these voters to be larger than the budget in the
        constructed priced control instance so that they cannot be deleted.}
        Otherwise, set $cost(v) = d_v - (\min\{d_{\mathcal{M}}(x_v,\ell)\ | \ \ell \in L\} - \epsilon)$.
    \item Let the set of unregistered voters $\widehat{W}$ consist of each voter
        $v \in V$ such that no $\ell \in L$ with
        $d_{\mathcal{M}}(x_v,\ell) \le d_v$ exists (i.e., they are not within their distance-bound
        to any polling place).
        Let $cost(v) = \min\{d_{\mathcal{M}}(x_v,\ell) \ | \ \ell \in L\} - d_v$.
\end{itemize}

For each $cost(v)$, we look for a rational number that is in the range $(cost(v), cost(v)+\frac{\epsilon}{|V|})$, so all parameter for the \citet{mia-fal:j:priced-control} algorithm are rational. Note that the maximal amount we added to the costs with $\frac{\epsilon}{|V|}$ is $\epsilon$, and thus if the cost could be met by $k$ it could be met by $\hat{k}$. If the cost could not be met by $k$, neither could it be by $\hat{k}$, since even giving all the added amount to a single agent would not change their ability to vote (since even $2\epsilon$ would not be enough).

Changing the distance-bound of a voter so that they can no longer vote corresponds to meeting the price to delete a registered voter, and changing the distance-bound of a voter so that they can vote, when they previously were not within their distance-bound to a polling place corresponds to adding a voter from the set of unregistered voters.

The reduction itself is straightforward: If the distance-bound of the voters in the geographic setting can be changed
using a total budget of $k$ (allowing them to vote/refrain from voting as needed) to ensure that $p$ is a winner,
then in the constructed instance of priced control, this corresponds to paying the same voters to be added/deleted
with a budget of $\hat{k}$ to ensure that $p$ is a winner, and vice versa.~\end{proofs}

More natural control actions to consider in future work are integrating control
by selecting polling places into the framework of multimode electoral control
introduced by \citet{fal-hem-hem:j:multimode}.
Briefly, multimode control considers the scenario where the election chair uses multiple
control actions to achieve their goal. One example is the control by adding voters
and deleting voters mentioned above for the priced setting. It is natural to consider
that an election chair that is able to select polling places would also be able to
control the structure of the election in other ways.

\section{Discussion and Future Work}

We introduced a new model for studying election problems that takes into account
geographic information. We examined the complexity of electoral control
by selecting voting locations in this geographic setting. We compared how the
complexity is affected by different parameters such as the two-party setting and
settings with an unbounded number of candidates,
and the cases where voters and polling places are placed on the line vs.\
being placed on the plane. Furthermore, we linked this setting with some of the existing problems in voting complexity in the nongeographic setting.

There are many different avenues for future work, as noted at the end of the previous sections. A specific open problem for future work is the complexity of polling place control when voters
can vote at most at two locations. Beyond that, further exploring how existing problems in election manipulation and control change once the geographic element is added is and interesting idea, as well as introducing of new models to capture natural scenarios in this setting. A more radical direction is to re-interpret the distance-bounds $d_{v}$ as a statistical measure, indicating probability of voting at a location of particular distance. This allows for richer settings, and more complex problems and analysis, and combines more deeply with geographic (urban/rural) considerations.

\bigskip

\noindent
{\bf Acknowledgments:} The authors thank Edith Hemaspaandra and anonymous reviewers for helpful comments.

\end{document}